\documentclass{article}
\usepackage{eurosym}
\usepackage{amssymb}
\usepackage{amsmath}
\usepackage{cite}

\begin{document}

\title{A Systematic Analysis of the Properties of the Generalised
Painlev\'e--Ince Equation}
\author{Andronikos Paliathanasis\thanks{%
Email: anpaliat@phys.uoa.gr} \\
{\ }
{\ \textit{Institute of Systems Science, Durban University of Technology }}\\
{\ \textit{PO Box 1334, Durban 4000, Republic of South Africa}} \\
\and P.G.L. Leach \\
{\ \textit{Institute of Systems Science, Durban University of Technology }}\\
{\ \textit{PO Box 1334, Durban 4000, Republic of South Africa}} \\
{\ \textit{School of Mathematical Sciences, University of KwaZulu-Natal }}\\
{\ \textit{Durban, Republic of South Africa}}}
\maketitle

\begin{abstract}
We consider the generalized Painlev\'{e}--Ince equation,
\begin{equation*}
\ddot{x}+\alpha x\dot{x}+\beta x^{3}=0
\end{equation*}%
and we perform a detailed study in terms of symmetry analysis and of the
singularity analysis. When the free parameters are related as $\beta =\alpha
^{2}/9~$the given differential equation is maximally symmetric and
well-known that it pass the Painlev\'{e} test. For arbitrary parameters we
find that there exists only two Lie point symmetries which can be used to
reduce the differential equation into an algebraic equation. However, the
generalized Painlev\'{e}--Ince equation fails at the Painlev\'{e} test,
except if we apply the singularity analysis for the new second-order
differential equation which follows from the change of variable $x=1/y.$ We
conclude that the Painlev\'{e}--Ince equation is integrable is terms of Lie
symmetries and of the Painlev\'{e} test.
\end{abstract}

\section{Introduction}

The Painlev\'{e}--Ince Equation,
\begin{equation}
\ddot{x}+3x\dot{x}+x^{3}=0,  \label{1}
\end{equation}%
is well-known to have a number of interesting features. Firstly it has eight
Lie point symmetries \cite{Mahomed85} with the algebra $sl(3,R)$ in the
Mubarakzyanov Classification Scheme \cite{Morozov58, Mubarakzyanov63a,
Mubarakzyanov63b, Mubarakzyanov63c} and so is linearisable by means of a
point transformation. Secondly it passes the Painlev\'{e} Test in terms of
the procedure of the ARS algorithm \cite{Ablowitz 78 a, Ablowitz 80 a,
Ablowitz 80 b} although in a way which was regarded as unacceptable at the
time. In 1993 Lemmer \textit{et al} \cite{Lemmer 93 a} showed that the
singularity was a simple pole and that the resonances for $a=1$ were a
perfectly normal 1 and the generic $-1$ whereas for $a=2$ they were $-1$ and
an unexpected $-2$. This result was not acceptable to some workers in the
field, but a subsequent programme initiated by Mac Feix \cite{Feix97} gave
substance to the result and a fuller exposition is to be found in the paper
of Andriopoulos \textit{et al} \cite{Andriopoulos 06 a} in which the whole
question of positive and negative resonances was answered in terms of
regions in the complex plane centred on the singularity. Thirdly the Painlev%
\'{e}--Ince Equation is a member -- the second -- of the Riccati Hierarchy
\cite{Euler 07 a, Euler 09 a, Euler 09 b, Euler 11 a}
based on the recursion operator, $D+x$ with $D=\frac{d\,}{dt}$, applied to $%
\dot{x}+x^{2}$ as initial member \footnote{%
One could simply start at $x$.}.

The generalised Painlev\'{e}--Ince Equation is defined as
\begin{equation}
\ddot{x}+\alpha x\dot{x}+\beta x^{3}=0  \label{2}
\end{equation}%
in which $\alpha $ and $\beta $ are constants, by means of some ingenious
manipulations. In this paper we approach the question of integrability of (%
\ref{2}) by means of the systemic methods of analysis, namely the search for
Lie point symmetries and the determination of the existence of a Laurent
series about a singularity.

\section{Symmetry Analysis}

The Lie point symmetries of (\ref{2}) are easily calculated using the
Mathematica add-on, SYM \cite{Dimas 05 a, Dimas 06 a, Dimas 08 a,
Andriopoulos 09 a}. For general values of $\alpha $ and $\beta $ there are
two symmetries, namely,
\begin{eqnarray}
\Gamma _{1} &=&\partial _{t}\quad \mbox{\rm and}  \label{3} \\
\Gamma _{2} &=&t\partial _{t}-x\partial _{x}  \label{4}
\end{eqnarray}%
except when the parameters are related according to
\begin{equation}
\beta =\alpha ^{2}/9.  \label{5}
\end{equation}%
Then there are the eight symmetries of the Painlev\'{e}--Ince Equation up to
the effect of the rescaling by $\alpha $.

The invariants for the symmetry, $\Gamma_1$ are
\begin{eqnarray}
&& u = x \quad \mbox{\rm and}  \label{6} \\
&& v = \dot{x}.  \label{7}
\end{eqnarray}
In the new variables (\ref{2}) becomes
\begin{equation}
vv^{\prime 3 }= 0 .  \label{8}
\end{equation}
In the new variables $\Gamma_2$ becomes $u\partial_u + v\partial_v$ when the
superfluous term in $t$ is ignored. The invariants for the once extended
form of this symmetry are
\begin{eqnarray}
&& w = \frac{v}{u^2} \quad \mbox{\rm and}  \label{9} \\
&& z = \frac{v^{\prime }}{u}.  \label{10}
\end{eqnarray}
With the invariants (\ref{9}) and (\ref{10} the first-order equation (\ref{8}%
) becomes the algebraic equation,
\begin{equation}
wz + \alpha w + q = 0.  \label{11}
\end{equation}

\section{Singularity Analysis}

We now turn to the analysis of (\ref{2}) in terms of the Painlev\'{e}-Test
as summarized in the ARS algorithm. The first step is to determine whether a
singularity exists and, if so, to calculate its coefficient. To this end we
make the substitution
\begin{equation}
x=a\tau ^{r}  \label{12}
\end{equation}%
into (\ref{2}), where $\tau =t-t_{0}$ and $t_{0}$ is the location of the
putative singularity. We find that the terms balance in exponent when $r=-1$%
, \textit{ie} the singularity is a simple pole. Moreover all terms are
dominant. When we replace $r$ with $-1$, the equation for the leading-order
coefficient is
\begin{equation}
\frac{2a}{\tau ^{3}}-\frac{\alpha a^{2}}{\tau ^{3}}+\frac{\beta a^{3}}{\tau
^{3}}=0  \label{13}
\end{equation}%
which has the solutions
\begin{equation}
a=0,\,\frac{\alpha -sqrt{\alpha ^{2}-8\beta }}{2\beta }\quad \mbox{\rm and}\,%
\frac{\alpha +sqrt{\alpha ^{2}-8\beta }}{2\beta }.  \label{14}
\end{equation}%
The first solution must be rejected as being irrelevant. The other two can
take any value as we are working in the complex time plane.

The second step of the ARS algorithm is the location of the resonances, that
is the exponents at which the remaining constants of integration enter the
Laurent expansion. We make the substitution
\begin{equation}
x = a\tau^{-1} + m\tau^{-1+s}  \label{15}
\end{equation}
into (\ref{2}) (recall that all terms in (\ref{2}) are dominant) and collect
the terms linear in $m$ for these are the terms at which a new constant
enters into the expansion. For that constant to be arbitrary the coefficient
of $m$ must be zero. The coefficient is a polynomial in $s$ and the values
of $s$ which render it zero are the resonances. When we do this, we obtain
\begin{equation}
s = -1,\, \frac{\alpha\sqrt{\alpha^2- 8\beta}-\alpha^2+8\beta}{2\beta}
\label{16}
\end{equation}
in the former case and
\begin{equation}
s = -1,\, \frac{-\alpha\sqrt{\alpha^2- 8\beta}-\alpha^2+8\beta}{2\beta}
\label{17}
\end{equation}
in the latter case. The value $-1$ is generic.

Acceptable values for the nongeneric resonance must be real and rational
with the practical limitation that fractional resonances should not have
denominators which are large for means that the complex plane is divided
into unworkable sections because of too many branch cuts. Note that there
are many possibilities for the second (nongeneric) resonance to be negative.
Consequently the `impossible' result for the Painlev\'{e}--Ince Equation
found by Lemmer \textit{et al} could be regarded as commonplace. Naturally
there are also a multitude of values for which the second (nongeneric)
resonance is irrational and/or complex.

\section{Conclusion}

We have seen that the generalised Painlev\'{e}--Ince Equation possesses two
Lie point symmetries and is reducible to an algebraic equation for all
values of the parameters. From singularity analysis we further see that the
singularity is always a simple pole. Various possibilities exist for the
nature of the Laurent expansion about this simple pole. It can be either a
Right -- expansion over a disc centered on the pole -- or a Left --
expansion about the pole over the complex plane without a disc -- Painlev%
\'{e} Series depending upon the values of the parameters. Alternatively it
can be a mess, thereby indicating nonintergability.

An interesting feature occurs if one inverts the dependent variable by
setting
\begin{equation}
x(t)=\frac{1}{y(t)}.  \label{18}
\end{equation}%
Then equation (\ref{2}) takes the following form%
\begin{equation}
y\ddot{y}-2\dot{y}^{2}+\alpha \dot{y}-\beta =0  \label{19}
\end{equation}
The two symmetries look the same under the coordinate transformation, except
that the plus sign becomes a minus sign in $\Gamma _{2}$. Naturally we are
not considering the particular case in which $\alpha ^{2}=9\beta .$

In terms of the singularity analysis for equation (\ref{19}) only the first
two terms are dominant, the singularity is a simple pole and the coefficient
of the leading-order term is unspecified. (A zeroth-order possibility also
exists, but that is not a singularity.) The resonances are at $-1$ and zero
in line with the arbitrary coefficient of the leading-order term. The
Laurent expansion corresponding to the simple pole is a Right Painlev\'{e}
Series. The interesting thing is that the coefficients of the higher-order
terms in the expansion vanish for the special relationship between $\alpha $
and $\beta $ which give rise to equations of maximal symmetry, that is
equations of Painlev\'{e}--Ince form. Hence, we can infer that the
generalized Painlev\'{e}--Ince equation (\ref{2}) passes the Painlev\'{e}
test under the coordinate transformation (\ref{18}) if and only if $\alpha
^{2}=9\beta $.

That is not the first example of a differential equation which pass the
singularity test under a coordinate transformation. A recent discussion on
that property can be found in \cite{ref01}.

\subsubsection*{Acknowledgements}

PGLL thanks the Durban University of Technology, the University of KwaZulu-Natal
and the National Research Foundation of South Africa for support. This work
was undertaken while we enjoyed the gracious hospitality of Surananee
University o Technology, Thailand.

\end{document}